\documentstyle[preprint,aps,amstex]{revtex}
\renewcommand{\d}{{\rm d}}
\renewcommand{\i}{{\rm i}}
\renewcommand{\Re}{{\rm Re}}
\renewcommand{\Im}{{\rm Im}}
\newcommand{\sgn}{{\rm sgn}}

\begin{document} 
\title{Persistent current induced by magnetic impurities$^*$}
\author{ {\sc P.\ Schwab$^{**}$} and {\sc U.\ Eckern}\\
 {\em Institut f\"ur Physik, Universit\"at Augsburg}\\
 {\em D-86135 Augsburg, Germany}
}
\date{\today}
\maketitle
\begin{abstract}
We calculate the average persistent current in a normal conducting,
mesoscopic ring in the diffusive regime.
In the presence of magnetic impurities, a contribution to the persistent current
is identified,
which is related to fluctuations in the electron spin density.
Assuming a spin-flip scattering rate which is comparable to the
Thouless energy $E_c$ and low temperature, this
new contribution to the persistent current
is of the order  $I\sim E_c^2/(kT\phi_0)$, 
which is considerably larger than the persistent current
induced by the electron-electron interaction.
\end{abstract}

\vspace{2cm}

\begin{tabular}{llll}
PACS: &05.30.Fk &--& Fermion systems and electron gas \\
      &72.10.Fk &--& Scattering by point defects, dislocations, and other imperfections\\
      &         &  & including Kondo effect\\
      &71.25.Mg &--& Electron energy states in amorphous and glassy solids
\end{tabular}

\vfill
\noindent
$^*$ submitted to Z. Phys. B \\
$^{**}$ Electronic mail: Peter.Schwab@@physik.uni-augsburg.de
\noindent
\newpage

\section{Introduction}
At low temperature, the magnetic response of small rings has a component which 
changes periodically with the applied magnetic flux \cite{Levy,Webb91,Mohanty96,Benoit93}.
Although this phenomenon, i.e.\ the existence of a persistent current, was predicted
many years ago \cite{Hund38,Byers61,Buttiker83}, the magnitude of the effect is not well understood:
The experimentally determined persistent currents in metallic rings \cite{Levy,Webb91,Mohanty96} 
are much larger than theory 
\cite{Bouchiat89,Montambaux90,Cheung89,Schmid91,vOppen91,Altshuler91,Ambeg90,UE91,%
Kopietz93,UE92,UE93a,Smith92,Berkovits93,Mueller94,Yoshioka94,Giamarchi95,Ramin95,Peters96,UE96}. 
predicts.
For systems where
the electron motion is diffusive, i.e.\ the mean free path $l$ is much smaller than the
circumference of the ring $L$, important theoretical  
results for the average persistent current are the following:
For non-interacting electrons, the current is of the order $I\sim \Delta/\phi_0$, where
$\Delta $ is the mean level spacing at the Fermi energy, 
$\Delta=(2{\cal N}_0 {\cal V})^{-1}$,
with ${\cal N}_0$ the density of states per spin and ${\cal V}$ the volume, 
and $\phi_0=h/e$ is the flux
quantum \cite{Schmid91,vOppen91,Altshuler91}. 
This result applies for temperatures below the
Thouless energy, which is given by $E_c= \hbar D/L^2$, $D=v_F l/3$,
but large compared to the mean level spacing.
For interacting electrons, the Coulomb interaction contributes to the persistent
current \cite{Ambeg90,UE91}, namely $I\sim \mu^* E_c/ \phi_0$, where $\mu^*$ characterizes the strength of the
interaction. For the metallic rings in the experiments the 
inequalities $\Delta \ll kT, E_c$ were fulfilled.

The experimental results can be fitted almost perfectly with the
`interacting' theory of \cite{Ambeg90} if we choose $\mu^*= \pm 0.3$. 
However the precise value of the interaction constant is not known from theory.  
To first order in the screened Coulomb interaction, the interaction constant
is of the correct order compared to experiment, however,
a repulsive interaction scales downwards when higher order corrections are taken into account 
\cite{Altshuler81a,Altshuler85,UE91}.
For example, for copper, the material of which the average persistent current has been measured,
$\mu ^* $ has been estimated \cite{UE91} to be smaller 
than the bare value of $0.3$ by a factor of about five to ten.

Many recent studies of persistent currents were devoted to the understanding
of the interplay of disorder and electron-electron interaction 
\cite{Kopietz93,UE92,UE93a,Smith92,Berkovits93,Mueller94,Yoshioka94,Giamarchi95,Ramin95,Peters96}.

In the present article we study 
a new mechanism which may induce a persistent current, namely
magnetic impurities. 
This is a novel phenomenon: While the
sensitivity of the persistent current to magnetic scattering has been predicted 
\cite{UE91,UE93,Yoshioka93,PS96a}
in previous studies, we point out that 
these investigations started from the theory of persistent currents without
magnetic impurities, and introduced magnetic defects as an additional perturbation.
In such an approach, an enhancement of
the persistent current is never to be found.
 
In contrast, 
in our recent article where we concentrated on persistent current {\em fluctuations}, 
we already concluded
that there should exists a new contribution to the {\em average} persistent current 
induced by magnetic impurities.   
Let us recall the argument \cite{PS96a}:
In a diffusive ring, the stochastic fluctuations of the persistent current are of the order
$\langle (\delta I)^2 \rangle \sim (E_c/\phi_0)^2$.
Changes in the impurity potential lead to a variation of the persistent current with this order
of magnitude.
Considering magnetic impurities as a spin-dependent scattering potential,
which changes slowly
due to spin relaxation processes, one finds temporal current fluctuations
of the order
$\langle (\delta I)^2 \rangle \sim (E_c/\phi_0)^2$
on the time scale of the spin relaxation time $\tau_K$.
Using the fluctuation-dissipation theorem we relate these current
fluctuations to the dynamic linear response $\chi(\omega)$ to a time-dependent 
magnetic flux $\phi(\omega)$ with the result
$\Im \chi(\omega\sim 1/\tau_K) \sim (E_c/\phi_0)^2 (\hbar/\tau_K)/k T$.
This suggests that at low frequency, 
there is also a contribution to the real part of $\chi$ 
of the order
$\chi(\omega \to 0) \sim (E_c/\phi_0)^2/kT$. In the static limit $\chi$ is
the derivative of the persistent current with respect to the flux:
$\chi(\omega=0)= \partial_\phi I(\phi)$.
In the present paper, we 
confirm this earlier suggestion \cite{PS96a} explicitly, namely
$I\sim E_c^2/(kT\phi_0)$.

In section II, we start with a few simple considerations, 
which demonstrate that a flux-dependence 
of the free energy induced by magnetic impurities may indeed exist.
Then we extend the calculations and discuss 
(1) the role of spin-orbit scattering,
(2) strong spin-flip scattering due to high concentrations of magnetic impurities, and
(3) strong spin-flip scattering due the Kondo effect.
\section{Weak spin scattering}

We start from the free energy of a spin free spin ($S=1/2$), 
given by
\begin{equation}
\label{eq1}
\Omega = - k T \ln \left( 2 \cosh{\mu_B H \over k T} \right)
\end{equation}
The spin is coupled to the conduction electrons; we assume the usual local exchange
Hamiltonian $H= J{\bf s}({\bf R})\cdot {\bf S}$,
where ${\bf s}({\bf R})$ is the spin density of the conduction electrons at the impurity site.
To first order in $J$, this coupling is equivalent to an additional magnetic field for the
impurity spin, and thus leads in Eq.\ (\ref{eq1}) to the replacement 
$2\mu_B H \to 2\mu_B H + J s^z({\bf R})$.
For electrons on a ring, the local spin density depends on the magnetic flux $\phi$ 
penetrating the ring. As a consequence the free energy is flux-dependent, and 
a persistent current $I= -\partial_\phi \Omega$ exists.

For the explicit calculation, we decompose the electron spin density in its flux-independent
mean value plus the flux-dependent fluctuations, 
$s^z({\bf R}) = \langle s^z({\bf R} )\rangle + \delta s^z({\bf R}, \phi)$. 
Then we expand $\Omega$ with respect to $\delta s^z$
and average over disorder, with the result
\begin{equation}\label{eq2}
\langle \Omega \rangle = \Omega ( \langle s^z \rangle ) - {1\over 8 \mu_B ^2} \chi^{zz}J^2
\langle \delta s^z({\bf R}) \delta s^z({\bf R}) \rangle
+ \cdots,
\end{equation} 
where $\chi^{zz} = - \partial_H^2 \Omega$ is the (longitudinal) susceptibility of the impurity.
These rather simple considerations already lead to a novel contribution to the persistent current.

Equation (\ref{eq2}) is valid for an arbitrary geometry. 
We consider a ring of circumference $L$ and transverse dimension
$L_\perp \ll L$. We assume diffusive motion of the electrons, i.e.\ the (elastic) mean free path
is much smaller than the circumference of the ring, $l\ll L$.

We evaluate the fluctuations of the local spin-density using the standard Green's functions technique. 
We average over impurity configurations, keeping only the diagrams with one particle-particle ladder,
i.e. one cooperon, as shown in Fig.\ \ref{fig1}.  
The flux sensitivity is determined by the long-wavelength, low-frequency contribution of the cooperon.
The result is ($\hbar = k = 1$)
\begin{eqnarray}
\langle \delta s^z({\bf R}) \delta s^z({\bf  R})  \rangle &=&
\left( {1\over \cal{ V} }\right)^3 T^2 \sum_{\epsilon, \omega} \sum_{\bf  k, k', q}
\sum_{s,s'} \sigma^z_{ss}\sigma^z_{s's'}C_{ss'ss'}({\bf q}, \omega, \epsilon)\nonumber\\ 
\label{eq3}& & \times G_s(\i \epsilon, {\bf k} )G_{s'}(\i (\epsilon -\omega), -{\bf k}+{\bf q} )    
G_s(\i \epsilon, {\bf k}' )G_{s'}(\i (\epsilon - \omega), -{\bf k}'+{\bf q} )
,\end{eqnarray}  
where $\cal V$ is the volume and $G_s$ the averaged Green's function for electrons with spin $s$, 
and $\epsilon$ and $\omega$ are odd and even Matsubara frequencies.
The relevant components of the cooperon, including the Zeeman effect, 
are given by
\begin{eqnarray}
C_{++++}= C_{----}  &=& {1\over 2\pi{\cal N}_0 \tau^2} {1\over |\omega| +Dq^2} \\
C_{+-+-}=C_{-+-+}^* &=& {1\over 2\pi{\cal N}_0 \tau^2}
{1\over |\omega| +Dq^2 + \i\omega_s \sgn(\omega)}
,\end{eqnarray}
provided $\epsilon(\epsilon -\omega) < 0$;
otherwise no singular contribution arises.
As usual, $\omega_s = 2\mu_B H$ denotes the Zeeman splitting;
see also App.\ A, where we summarize the results for the cooperon including 
spin-flip and spin-orbit scattering.
For simplicity, we consider the limit $1/\tau_s=1/\tau_{so}=0$ for the moment.
The diffusion constant, $D$, is given by $D=v_F^2 \tau/3$, where $\tau$ denotes the elastic scattering time.

Equation (\ref{eq3}) is similar in form to the expression for the fluctuations of the local
electron density, which have been evaluated in \cite{Ambeg90} in order to 
determine the Coulomb interaction
contribution to the persistent current. In \cite{Ambeg90} both Hartree and
exchange contribution to the persistent current are calculated. The results presented here
and below correspond to the Hartree terms; we neglect the exchange 
contributions, which are insignificant in most cases. In order to evaluate Eq.\ (\ref{eq3})
we rely on the results given in \cite{Ambeg90}. 

For the ring geometry, the transverse dimensions of ${\bf q}$ are frozen out, 
such that it can be considered
one-dimensional, assuming the values $q= 2\pi(n + 2\phi/\phi_0)/L$, where $\phi_0=h/e$ is the
flux quantum and $n$ is an integer number.
The summations over ${\bf k}$ and ${\bf k'}$ are converted into integrals, and from each integration
we find a factor $(2\pi {\cal N}_0 \tau)$. The $\epsilon$ summation is then feasible 
since the $\epsilon$ dependence is only due to the condition that $\epsilon(\epsilon-\omega)<0$.
Finally we arrive at
\begin{equation}\label{eq6}
\langle \delta s^z({\bf R}) \delta s^z({\bf R}) \rangle = {4 {\cal N}_0 \over {\cal V}}
T \sum_{\omega >0 } \omega \sum_q 
\left( {1\over \omega + Dq^2}- \Re{1\over \omega +Dq^2 + \i \omega_s}  
\right)
\end{equation}
The spin fluctuations are periodic in $\phi$ with periodicity $h/2e$, and
can thus be represented as a Fourier series,
\begin{equation} \label{eq6a}
\langle \delta s^z({\bf R}) \delta s^z({\bf R}) \rangle = 
\sum_{m= -\infty}^{\infty} A_m \exp( 2\pi \i m \phi/\phi_0)
.\end{equation}
Note that only the even components are non-zero,
with the result
\begin{equation}
A_{2m} (T, \omega_s) = {2{\cal N}_0 \over {\cal V} } T\sum_{\omega >0} \omega
\int_{-\infty}^{\infty} \d x e^{-2\pi \i m x}
\left(
{1\over \omega+4\pi^2 Dx^2/L^2 } - \Re {1\over \omega +4\pi^2 Dx^2/L^2 +\i \omega_s}
\right) 
.\end{equation}
All even coefficients are related to the second coefficient at 
rescaled temperature and Zeeman energy,
$A_{2m}(T,\omega_s) = A_2(m^2 T, m^2\omega_s)/|m|^3$,
so it is sufficient to calculate $A_2$.
Since $A_m=A_{-m}$ only terms proportional to $\cos( 2\pi m \phi/\phi_0)$ survive
in Eq.\ (\ref{eq6a}).
The Fourier series of the grand potential is
\begin{equation}
\langle \Omega \rangle = \langle \Omega_0 \rangle 
+ \sum_{m=1}^\infty \langle \Omega_m \rangle \cos (2\pi m \phi/\phi_0)
,\end{equation}
with 
$\langle \Omega_m \rangle = - \chi^{zz} J^2 A_m /4 \mu_B^2$.
Finally, the Fourier components of the persistent current are given by
$\langle I_{m}\rangle =  e m\langle \Omega_{m} \rangle $.

Numerical results for the first non-vanishing Fourier component of the grand potential, 
$\langle\Omega_2 \rangle$, as a function of temperature and for different values of the Zeeman
energy are shown in Fig.\ \ref{figDomegaSd}. 
For a finite but small concentration of magnetic
impurities, we have to multiply the result with the number of impurities $N_s$. 
In Fig.\ \ref{figDomegaSd} 
we choose the concentration such that the spin-flip scattering rate equals the Thouless energy:
$1/\tau_s = 2\pi n_s {\cal N}_0 J^2 S(S+1)= E_c$.
We find that
$\langle \Omega_2 \rangle$ goes to zero both in the limit of a weak and a strong magnetic field:
For a weak magnetic field, there is no electron spin polarization and consequently no 
fluctuations in the spin polarization, while for a
strong magnetic field ($\omega_s \gg T $), 
$\langle \Omega_2 \rangle$ goes to zero since the susceptibility
goes to zero. The maximum value is found for
intermediate values of the magnetic field, where there are fluctuations in the 
electron spin density but
the impurity spins are not yet fully polarized, such that $\chi^{zz}\ne 0$.
\section{Strong spin scattering}
In the case of strong spin-orbit scattering or strong spin-flip scattering, there are fluctuations in 
the local electron spin density even without Zeeman effects and
we find a significantly enhanced persistent current.

\subsection{Strong spin-orbit scattering}
Let us consider spin-orbit scattering, assuming that spin-flip scattering is weak.
In this case it is straightforward to generalize the calculations described above. 
We take into account that there exist fluctuations in the electron spin density
not only in $z$-direction, but also in $x$- and $y$-direction.
Equation (\ref{eq2}) generalizes to
\begin{equation} \label{eq8}
\langle \Omega \rangle  =
\Omega( \langle s^z \rangle )  
- {1\over 8 \mu_B^2} J^2 \left(
\chi^{zz} \langle \delta s^z \delta s^z \rangle + 
4 \chi^{+-}
\langle \delta s^-\delta s^+ \rangle 
\right)
,\end{equation}
with the susceptibilities given by
\begin{eqnarray}
{1\over 4 \mu_B^2} \chi^{zz} &=& {1\over 4T} \left(1-\tanh^2 {\omega_s\over 2T} \right) \\
{1\over 4 \mu_B^2} \chi^{+-} &=& {1\over 2\omega_s} \tanh{\omega_s\over 2T}
.\end{eqnarray}
The fluctuations in the local spin density are 
\begin{eqnarray}
\langle \delta s^z \delta s^z \rangle &=&
{2T\over {\cal V}} \sum_{\omega>0} {\omega\over 2\pi} (2\pi{\cal N}_0\tau)^2 \sum_{q,s,s'}
\sigma^z_{ss}\sigma^z_{s's'}C_{ss'ss'}\\
\langle \delta s^+ \delta s^- \rangle &=&
{2T\over {\cal V}} \sum_{\omega>0} {\omega\over 2\pi} (2\pi{\cal N}_0\tau)^2 \sum_{q} C_{-++-}
.\end{eqnarray}
with the expressions for the cooperon given in App.\ A. 

In the limit of zero spin-orbit scattering, 
$C_{-++-}$ is zero and thus the transverse spin fluctuations are zero.
Then the longitudinal fluctuations are identical to the expression given in Eq.\ (\ref{eq6}). 
For very strong spin-orbit scattering,
on the other hand, ($1/\tau_{so} \gg E_c$ and $\omega_s^2 \ll E_c /\tau_{so}$),
the cooperon simplifies considerably:
$C_{++++} \approx 0$, $2\pi{\cal N}_0 \tau^2 C_{+-+-} \approx 1/2N_0$, 
and $2\pi{\cal N}_0 \tau^2 C_{+--+}\approx -1/2N_0$, where $N_0=|\omega|+ Dq^2$.
The flux-dependent part of the thermodynamic potential is then given by
\begin{multline} \label{eq13} 
- {N_s \over 8 \mu_B^2} J^2 \left(
\chi^{zz} \langle \delta s^z \delta s^z \rangle +
4 \chi^{+-}
\langle \delta s^-\delta s^+ \rangle
\right)=\\
 {1\over \tau_s}
\left[ {1\over 3} \left( 1-\tanh^2{\omega_s\over 2T}\right) + {4T\over 3\omega_s}\tanh{\omega_s\over 2T} \right]
\sum_{q,\omega>0}{ \omega \over 2\pi}{1\over \omega+Dq^2}
.\end{multline}

Fig.\ \ref{figSpinOrb} shows the amplitude of the second harmonic, $\langle \Omega_2 \rangle $,
for the same values of $\omega_s$, $T$, and $1/\tau_s$ as in Fig.\ \ref{figDomegaSd}. 
Spin-orbit scattering changes the sign of 
$\langle\Omega_2\rangle $, it does not go to zero for 
$\omega_s\to 0$, and especially the amplitude of 
$\langle \Omega_2\rangle$ 
is much larger than without spin-orbit scattering.
For low temperature, the flux-dependence due to magnetic impurities
dominates the flux-dependence due to the Coulomb interaction. The latter
for low temperature is given by \cite{Ambeg90}
$\langle \Omega_2 \rangle =(4/\pi) \mu^* E_c \exp( -T/3E_c )$. In the figure we use the
estimated value for the effective interaction, $\mu^* \approx 0.3/5 =0.06$.

We note that for the temperatures in the figure, 
the flux-dependence of the grand potential due to magnetic impurities is well approximated by
\begin{equation}
\langle\Omega_{2}\rangle = {1\over \tau_s} {1\over T}
\left( 1-\tanh^2{\omega_s\over 2T} +{4T\over \omega_s}\tanh{{\omega_s\over 2T }}\right)
{E_c \over 3\pi^2} e^{-T/ 3E_c}      
.\end{equation}
The exponential law, $\exp(-T/3E_c)$, reflects the decay of the 
flux-dependence of the spin fluctuations,
while the other factors are due to the temperature dependence of the impurity susceptibility.
At low temperature ($T \ll E_c$), and weak magnetic field ($\omega_s \ll  T$), 
$\langle \Omega_2 \rangle$ is thus proportional to the inverse temperature, 
\begin{equation}\label{eq15}
\langle \Omega_2 \rangle = {1\over \pi^2} {1\over \tau_s} {1\over T} E_c
.\end{equation} 
The $1/T$-divergence is cut off for $T \ll \omega_s $, where we find
\begin{equation}\label{eq16} 
\langle \Omega_2 \rangle = {4\over 3\pi^2}{1\over \tau_s}{1\over \omega_s} E_c
.\end{equation}
\subsection{Strong spin-flip scattering -- high concentration of impurities}
It is well known that the cooperon is sensitive to magnetic scattering. 
As a consequence, Eqs.\ (\ref{eq2}) and (\ref{eq8}) are only applicable in the 
limit of weak spin-flip scattering, i.e.\ for a low concentration of magnetic impurities.
However, simply inserting the cooperon including spin-flip scattering
leads to incorrect results, since when 
calculating the thermodynamic potential within
the diagrammatic approach, special care has to be taken with symmetry factors.

A typical diagram for the higher order corrections is shown in Fig.\ \ref{figHighord}.
The curly line represents scattering at magnetic impurities.
We consider $\omega_s=0$. In this case, the curly line
does not transfer energy, since the thermal average of two spin-operators is time independent,
e.g.\ $\langle S^z(\tau)S^z(0)\rangle =S(S+1)/3$.
The summation over all contributions of the type shown in Fig.\ \ref{figHighord} leads to 
\begin{eqnarray}\label{eq18}
\langle \Omega \rangle &=& -  \sum_{q,\omega} {|\omega |\over 4\pi}
 \sum_{n=1}^\infty  {1\over n} 
\left( -1  /\tau_s \over |\omega |+Dq^2 +1/\tau_s \right)^n \nonumber\\
&& - 3\sum_{q,\omega} {|\omega|\over 4\pi} \sum_{n=1}^\infty   {1\over n} 
 \left( 1/ 3\tau_s \over |\omega|+Dq^2 +1/\tau_s +4/3\tau_{so}\right)^n
.\end{eqnarray}
Performing the $n$-summation, we find
\begin{equation}\label{eq19}
\langle \Omega \rangle = {1\over 2} \sum_{q,\omega} {|\omega|\over 2\pi} \ln \left[
\left( {|\omega|+ Dq^2 +2/\tau_s \over |\omega| + Dq^2+ 1/\tau_s } \right) \cdot 
\left( {|\omega|+ Dq^2 +2/3\tau_s +4/3\tau_{so} \over |\omega|+ Dq^2 +1/\tau_s+4/3\tau_{so}}
\right)^3\right]
.\end{equation}
An expression of this form has been found for the persistent current {\em fluctuations}
\cite{UE93,PS96a}, and we use results given there in order to evaluate the Fourier
expansion of the {\em average} persistent current.
We find
\begin{equation}\label{eq20}
T\sum_{q, \omega >0} {|\omega |\over 2\pi} \ln(|\omega |+Dq^2+ \gamma) =
-{6\over \pi^2 }E_c^2 \sum_{m=1}^\infty {h^0_m(\gamma)\over m^5 }\cos( 4\pi m \phi/\phi_0)
;\end{equation}
for low temperature ($T \ll E_c, \gamma$), 
\begin{equation}
h_m^0(\gamma) = e^{-\sqrt{ \gamma m^2/E_c}} \left( 
1+ \sqrt{\gamma m^2 / E_c} + { \gamma m^2\over 3  E_c} \right)
.\end{equation}
Using these relations, we can expand Eq.\ (\ref{eq19}) as follows: 
\begin{equation}\label{eq22}
\langle \Omega\rangle = -{1\over T} {6\over \pi^2} E_c^2  \sum_{m=1}^\infty {h_m\over m^5} \cos (4\pi m \phi/\phi_0)
,\end{equation}
with
\begin{equation}\label{eq23a}
h_m = h_m^0(2/\tau_s) +3h_m^0(2/3\tau_s+4/3\tau_{so} ) - 
      h_m^0(1/\tau_s) -3h_m^0(1/\tau_s+4/3\tau_{so})
.\end{equation}
Flux-independent terms have been dropped again.
Note that in Eq.\ (\ref{eq22})   
the prefactor $1/T$ appears which is due to the impurity susceptibility, 
resulting in rather large Fourier components at low temperature. 
Expanding Eq.\ ({\ref{eq23a}) in the limit of weak spin-flip scattering, 
but strong spin-orbit scattering, 
we find
$h_1=-1/6\tau_s E_c$,
recovering the result given in Eq.\ (\ref{eq15}) for the grand potential.
For vanishing spin-orbit scattering and weak spin-flip scattering, the leading term is quadratic in the
spin flip scattering rate:  
$h_1=(1/\tau_s E_c)^2/18$. This agrees with our result that 
the contribution to $\langle \Omega_2 \rangle$ which is linear in $1/\tau_s$ 
vanishes in the limit $\omega_s \to 0$.
   
For further illustration, we plot the second harmonic of the grand potential 
as a function of the spin-flip scattering rate in Fig.\ \ref{figHighord}.
The absolute values of 
$\langle \Omega_2 \rangle$ are comparable in size for weak and strong spin-orbit scattering.
Over a large range of the spin-flip scattering rate, we roughly have 
$|\langle \Omega_2 \rangle| \sim 0.1 E_c^2/T$.
Without spin-orbit scattering, the persistent current is diamagnetic for small magnetic flux, whereas
the persistent current is paramagnetic in the case of strong spin-orbit scattering.

In the limit of very strong spin-flip scattering, the persistent current approaches zero.
Without spin-orbit scattering, the dominant term in Eq.\ (\ref{eq23a}) is
$h_m \approx 3 h_m^0(2/3\tau_s)$. In the limit of strong spin-orbit scattering,
$h_m \approx -h_m^0(1/\tau_s)$ is dominant. Thus the persistent current, as a function of the spin-flip
scattering rate, approaches zero much faster for strong spin-orbit scattering than without spin-orbit
scattering, as is apparent in the figure.
\subsection{Strong spin-flip scattering -- the Kondo effect}
Perturbation theory in the coupling constant $J$, as in Eqs.\ (\ref{eq2}) and ({\ref{eq8}),
only works well far above the Kondo temperature, which is given by
$T_K \sim \epsilon_F \exp(-1/2|J|{\cal N}_0)$.
In this section, we describe the magnetic impurities in the framework of the
single impurity Anderson model, and calculate the grand potential to first order
in the on-site interaction $U$. In averaging over impurities, we again only keep diagrams with
one cooperon. Following the ideas of Hewson's renormalized perturbation expansion \cite{Hewson93},
we introduce the renormalized Green's functions and a renormalized interaction.
This allows calculation of the persistent current for temperatures far below the Kondo temperature.
We consider low concentration of paramagnetic impurities, and $\omega_s= 1/\tau_{so}=0$.

The contribution to the grand potential to first order in the Coulomb repulsion between
$d$-electrons is given by
\begin{equation} \label{eq23}
U n_{d\uparrow}n_{d\downarrow} = U T^2 \sum_{\epsilon_1\epsilon_2} 
G_{d\uparrow}(\i \epsilon_1)G_{d\downarrow}(\i\epsilon_2)
.\end{equation}
Here, $G_d$ is the Green's function of $d$-electrons of the 
non-interacting Anderson model. 
The $d$-electrons are coupled to conduction electrons on a disordered ring. 
The $d$-Green's function averaged over disorder in the
conduction band is given by
\begin{eqnarray} \label{eq24}
G_{d\sigma}(\i \omega)&=&\left[ 
\i \omega -\epsilon_d+\mu + |V|^2 \int {\d^3 k\over (2\pi)^3} G_\sigma(\i \omega, {\bf k}) \right]^{-1}
\\
&=& {1\over \i \omega -\epsilon_d +\mu + \i \delta \sgn (\omega)}
,\end{eqnarray}
where $V$ is the hybridization of the $d$-electrons with the conduction band,
and $\delta$ the broadening of the $d$-level due to hybridization.
Note that the averaged $d$-Green's function in this approximation does not depend on disorder. 
However the average of the product of two Green's functions also involves  
fluctuations; in averaging Eq.\ (\ref{eq23}), 
we keep the diagrams shown in Fig.\ \ref{figAnderson}, with the result
\begin{equation} \label{eq26}
\langle \Omega(\phi) \rangle = {V^4 U T^2\over {\cal V}} \sum_{\epsilon, \omega}
\left( G_{d\uparrow}(\i\epsilon)\right)^2
\left( G_{d\downarrow}(\i\omega -\i \epsilon) )\right)^2
\sum_q (2\pi{\cal N}_0 \tau)^2 C_{+-+-}(q,\omega,\epsilon)
.\end{equation}
The flux-dependence of this equation is due to the low energy singularity of the cooperon.
Flux independent contributions to the grand potential have been dropped.
In order to calculate the flux-dependence of $\Omega$ for temperatures  
below the Kondo temperature,
we generalize this equation replacing $G_{d\sigma}(\i\epsilon)$ with the exact
Green's function $\tilde G_{d\sigma}(\i \epsilon_n)$ of the clean Anderson model, and
replacing $U$ with the exact irreducible, antisymmetrized four-point vertex function.
This leads to
\begin{eqnarray} 
\langle \Omega(\phi) \rangle &=& {V^4 T^2 \over {\cal V}} \sum_{\epsilon, \omega}
\left( \tilde G_{d\uparrow}(\i\epsilon)\right)^2
\left( \tilde G_{d\downarrow}(\i\omega-\i\epsilon)\right)^2 \nonumber\\
\label{eq27}& &\times
\Gamma_{\uparrow\downarrow}^{\uparrow\downarrow}(\i \epsilon,\i\epsilon , \i\omega-\i\epsilon, \i\omega -\i\epsilon)
\sum_q (2\pi{\cal N}_0 \tau)^2 C_{+-+-}(q, \omega, \epsilon)
.\end{eqnarray}
In the limit where the Kondo temperature is large compared to the Thouless energy and temperature, we
can evaluate Eq.\ (\ref{eq27}) explicitly, since 
for low energy and temperature, analytic expressions for the $d$-Green's function and the vertex are known \cite{Hewson93,Hewson93a}:
\begin{eqnarray}
\tilde G_{d\sigma}(\i\epsilon )&\approx &
{z\over \i\epsilon -\tilde \epsilon_d + \mu + \i \tilde \delta} \\ 
\Gamma_{\uparrow\downarrow}^{\uparrow\downarrow}(0,0,0,0) &\approx & \tilde U
\end{eqnarray}
with
\begin{equation}
\tilde \epsilon_d = \mu, \quad \tilde \delta= 4T_K /\pi w = \pi {\cal N}_0 V^2 z, \quad
\tilde U = \pi \tilde \delta /z^2 ;\end{equation}
$w\approx 0.41$ is the Wilson number.
The flux-dependence of Eq.\ (\ref{eq27}) arises from energies of the order of the Thouless
energy, so for $T_K \gg E_c, T$,
we approximate the $d$-Green's functions and the vertex by the constant values 
which they assume for $\epsilon\to 0$.
The flux-dependent contribution to the grand potential is then given by
\begin{equation} \label{eq32}
\langle \Omega (\phi) \rangle
= {1\over {\cal N}_0 {\cal V}} {\pi w \over T_K}
T\sum_{q,\omega >0 } {\omega \over 2 \pi} {1\over \omega+ Dq^2}
.\end{equation}
This is of the same structure as the flux-dependence in the weak coupling limit, 
see Eqs.\ (\ref{eq8}) and (\ref{eq13}); 
thus $\langle \Omega (\phi) \rangle$ contains the following factors: 
\begin{enumerate}
\item The scattering amplitude of electrons with the impurities. For $T\ll T_K$ the scattering
amplitude is given by the unitarity limit for $s$-wave scattering. The
explicit result for the scattering rate is $1/\tau_s = 2 N_s/ \pi {\cal N}_0 {\cal V}$. 
\item 
 The magnetic susceptibility of the impurity, which is proportional to the inverse Kondo
temperature, $\chi = \mu_B^2 w/ T_K$.
\item
A factor which describes the diffusive motion of the electrons around the ring.
\end{enumerate}
From Eq.\ (\ref{eq32}), we determine the Fourier components
of the grand potential ($T\to 0$)
\begin{equation}\label{eq33}
\langle \Omega_{2m} \rangle = N_s {2w\over \pi}{\Delta \over T_K} {E_c \over m^3}
,\end{equation}
where we assumed a finite number of impurities; 
recall that $\Delta = 1/2{\cal N}_0 {\cal V}$.

In the presence of spin-orbit scattering we determine the flux-dependence of the 
grand potential from a product of spin-dependent 
vertex function and cooperon of the form
\begin{equation}
{1\over 2} \sum_{\sigma_1\sigma_2\sigma_3\sigma_4} 
\Gamma^{\sigma_1\sigma_2}_{\sigma_3\sigma_4}C_{\sigma_1\sigma_2\sigma_3\sigma_4}
={\tilde U\over 2} \left( C_{+-+-}+C_{-+-+}-C_{-++-}-C_{+--+} \right)
.\end{equation}
In this combination of the components of the cooperon, which is the singlet component, spin-orbit scattering drops out.

\section{Discussion}
We showed that magnetic impurities contribute significantly to the average persistent current in 
mesoscopic rings. The persistent current is proportional to the impurity susceptibility.
Assuming a Curie law, the persistent current thus becomes large for low temperature
and dominates the Coulomb interaction contribution to the persistent current.
The persistent current induced by magnetic impurities is large for a wide range of parameters, i.e.\ 
in the presence or the absence of spin-orbit scattering and also for rather high concentrations of
impurities.

Even if the impurity spins are screened due to the Kondo effect, a contribution to 
the persistent current remains. 
For example for $ 10^4$ impurities with $T_K \sim 1$K  
we find from Eq.\ (\ref{eq33}) 
$\langle \Omega_2 \rangle \sim 0.5 E_c$, when we insert $\Delta = 0.2 $mK which has been reported in
the experiment \cite{Levy}. This is close to the experimental result, where 
$\langle \Omega_2 \rangle \approx 0.3 E_c$.
However, we have to mention that 
for large concentration of magnetic impurities there are deviations from the linear
dependence of the persistent current on the concentration of impurities
which is assumed in Eq.\ (\ref{eq33}). 

A precise calculation of the persistent current as a function of
concentration of magnetic impurities is not easy, since one has   
to take impurity-impurity interactions into account.
Especially, one needs the full energy dependence of the
$d$-Green's functions and of the vertex $\Gamma$.
In order to estimate the current, we approximate $G_d$ and
$\Gamma$ by constants, up to a cut-off energy of the order
of $T_K$. Thus the dimensionless effective electron-electron 
interaction due to magnetic impurities is given by  
\begin{eqnarray}
\nu & = & {\cal N}_0 N_s {|V|^4\over {\cal V}^2 }|G_d(\i\omega=0)|^4
\Gamma_{\uparrow\downarrow}^{\uparrow\downarrow}(0,0,0,0) \\
&=& N_s {\Delta w\over 2 T_K} = n_s {w \over 4{\cal N}_0 T_K}
, \end{eqnarray}
with $n_s=N_s/{\cal V}$ the concentration of impurities, and Eq.\ (\ref{eq33}) reads
\begin{equation}
\langle\Omega \rangle = 2 \nu T \sum_{q, \omega>0}{\omega \over \omega+Dq^2}
.\end{equation}
Including impurity self-energy diagrams in the calculation of the cooperon leads to a renormalization 
of the diffusion pole \cite{PS96a},
\begin{equation}
\omega+Dq^2  \rightarrow (1+\nu)\omega + Dq^2
,\end{equation}
and finally, higher order terms renormalize $\nu$ logarithmically
 \cite{Altshuler85,Altshuler85a,UE91},
\begin{equation}
\nu \rightarrow \nu^*= {\nu\over 1+[\nu /( 1+\nu)] \ln T_K/ E_c}
.\end{equation}
However, in any real metal there is the Coulomb interaction
and the electron-phonon interaction which complicate our situation further.
For low concentration of magnetic impurities, the contribution
to the persistent current due to Coulomb interaction and magnetic impurities
simply adds, but at a high concentration, this certainly is not the case.
In order to estimate the current, we approximate the interactions in the
conduction band introducing a second interaction constant $\mu$.
If the Coulomb interaction is dominant, $\mu$ is 
positive and the cut-off for the interaction is of the order of the
Fermi energy, $\omega_c \approx \epsilon_F$. 
If the phonons dominate, $\mu$ is negative with a cut-off
of the order of the Debye energy, $\omega_c \approx \omega_D$.
We assume that the relation $\omega_c \gg T_K \gg E_c$ holds.
Having two different interactions, we first renormalize $\mu$ down
to the Kondo temperature,
\begin{equation}
\mu \rightarrow \mu^* ={\mu\over 1+\mu \ln( \omega_c/T_K )}
,\end{equation}
and then the sum of both is scaled down to $E_c$, with the final
result
\begin{equation}
\langle \Omega \rangle =2 {\nu+\mu^*\over
1+[(\nu+\mu^*)/(1+\nu)] \ln(T_K /E_c) }
T\sum_{q, \omega>0} {\omega \over (1+\nu)\omega + Dq^2 }
.\end{equation}
The Fourier components of the grand potential (and of the persistent current) as 
a function of concentration of magnetic impurities are then given by ($T=0$)
\begin{equation}
\langle \Omega_{2m} \rangle
={\nu+\mu^* \over
   1+ [(\nu+\mu^*)/(1+\nu)] \ln(T_K /E_c) }{4\over \pi}{E_c\over m^3}{1\over (1+\nu)^2}
.\end{equation}
With the parameters used above, $\nu \approx 0.5$, and there are already strong
deviations from a linear dependence of $ \langle \Omega_{2m} \rangle$ on $\nu$ as given in
Eq.\ (\ref{eq33}). 
  
In conclusion we have shown that magnetic impurities contribute
significantly to the persistent current. Thus we predict that future experimental studies
of persistent currents in the presence of magnetic impurities promise 
to yield most interesting results -- and in particular a clear test of the theoretical concepts.
\newpage
\appendix
\section{Cooperon in the presence of spin-effects}
The spin dependence of the cooperon has been discussed at various places 
in the literature \cite{Hikami80,Altshuler85,Altshuler85a}.
We assume that the 
rate for scattering at non-magnetic impurities, denoted here for clarity by $1/\tau_0$,
is large compared to 
the spin-flip scattering rate, the spin-orbit scattering rate, and the Zeeman splitting, i.e.\
$1/\tau_0 \gg 1/\tau_s, 1/\tau_{so}, \omega_s$.
The components of the cooperon are determined from the equation
\begin{equation}
C_{\alpha\beta\gamma\delta}=C^0_{\alpha\beta\gamma\delta}+
C^0_{\alpha\beta\mu\nu}\Pi_{\mu\nu}C_{\mu\nu\gamma\delta}
,\end{equation}
see Fig.\ \ref{figCooperon};
summation over $\mu$ and $\nu$ is implied here. The bare cooperon is given by
\begin{equation}
C^0_{\alpha\beta\gamma\delta}= {1\over \tau_0} \delta_{\alpha\gamma}\delta_{\beta\delta}
+{1\over 3} \left( {1\over \tau_s} - {1\over \tau_{so}} \right)
\vec\sigma_{\alpha\gamma} \vec\sigma_{\beta\delta}
.\end{equation}
The quantity $\Pi_{\mu\nu}$ is determined from the product over two Green's functions,
integrated over momentum,
\begin{equation}
\Pi_{\mu\nu} = \int {\d^3 k \over (2\pi)^3} G_{\mu}(\i \epsilon , {\bf k} )
G_{\nu}(\i \epsilon- \i\omega, -{\bf k}+ {\bf q} )
,\end{equation}
where $\mu,\nu= \pm 1$ are spin indices. In the long-wavelength, low-frequency limit
the result is 
\begin{equation}
\Pi_{\mu\nu} = 2\pi {\cal N}_0 \tau \left[ 1 -\tau\left(
|\omega|+ Dq^2 +\i (\mu-\nu)\omega_s\sgn(\omega)/2 \right) \right] 
,\end{equation}
where $1/\tau={1/\tau_0}+1/\tau_s+1/\tau_{so}$.
For the summery of the results we define 
\begin{eqnarray}
N_0 &=& |\omega| + Dq^2 +2/\tau_s \\
N_1 &=& |\omega| + Dq^2 +2/3\tau_s + 4/3\tau_{so}
,\end{eqnarray}
to express the non-zero components of the cooperon in a concise way:
\begin{eqnarray}
2\pi {\cal N}_0 \tau^2 C_{++++} &=& 1/N_1 \\
2\pi {\cal N}_0 \tau^2 C_{+-+-} &=& [N_0+N_1 -2\i \omega_s \sgn (\omega)]/2{\cal D}\\
2\pi {\cal N}_0 \tau^2 C_{+--+} &=& (N_0-N_1)/2{\cal D} 
\end{eqnarray}
with ${\cal D}=N_0N_1 + \omega_s^2$,
and
\begin{equation}
C_{-+-+}=C_{+-+-}^*, \quad C_{-++-}=C_{+--+}, \quad C_{----}=C_{++++}^* 
.\end{equation}
%

\begin{thebibliography}{10}

\bibitem{Levy}
L.~P. L\'evy, G. Dolan, J. Dunsmuir, and H. Bouchiat, Phys. Rev. Lett. {\bf
  64},  2074  (1990);
L.~P. L\'evy, Physica B {\bf 169},  245  (1991).

\bibitem{Webb91}
V. Chandrasekhar {\it et~al.}, Phys. Rev. Lett. {\bf 67},  3578  (1991).

\bibitem{Mohanty96}
P. Mohanty, E.~M.~Q. Jariwala, M.~B. Ketchen, and R.~A. Webb,  in {\em Quantum
  Coherence and Decoherence}, edited by K. Fujikawa and Y.~A. Ono 
  (Elsevier, Amsterdam, 1996), p. 191.

\bibitem{Benoit93}
D. Mailly, C. Chapelier, and A. Benoit, Phys. Rev. Lett. {\bf 70},  2020
  (1993).

\bibitem{Hund38}
F. Hund, Ann. Phys. (Leipzig) {\bf 32},  102  (1938), ibid. {\bf 5}, 1 (1996) (English 
translation).

\bibitem{Byers61}
N. Byers and C.~N. Yang, Phys. Rev. Lett. {\bf 7},  46  (1961).

\bibitem{Buttiker83}
M. B\"uttiker, Y. Imry, and R. Landauer, Phys. Lett. {\bf 96A},  365  (1983).

\bibitem{Bouchiat89}
H. Bouchiat and G. Montambaux, J. Phys. (Paris) {\bf 50},  2695  (1989).

\bibitem{Montambaux90}
G. Montambaux, H. Bouchiat, D. Sigetti, and R. Friesner, Phys. Rev. B {\bf 42},
   7647  (1990).

\bibitem{Cheung89}
H.-F. Cheung, E.~K. Riedel, and Y. Gefen, Phys. Rev. Lett. {\bf 62},  587
  (1989).

\bibitem{Schmid91}
A. Schmid, Phys. Rev. Lett. {\bf 66},  80  (1991).

\bibitem{vOppen91}
F. von Oppen and E. Riedel, Phys. Rev. Lett. {\bf 66},  84  (1991).

\bibitem{Altshuler91}
B.~L. Altshuler, Y. Gefen, and Y. Imry, Phys. Rev. Lett. {\bf 66},  88  (1991).

\bibitem{Ambeg90}
V. Ambegaokar and U. Eckern, Phys. Rev. Lett. {\bf 65},  381  (1990).

\bibitem{UE91}
U. Eckern, Z. Phys. B {\bf 82},  393  (1991).

\bibitem{Kopietz93}
P. Kopietz, Phys. Rev. Lett. {\bf 70},  3123  (1993), see also: G.\ Vignale,
  ibid. {\bf 72}, 433 (1994); A.\ Altland and Y.\ Gefen, ibid. {\bf 72}, 2973
  (1994).

\bibitem{UE92}
U. Eckern and A. Schmid, Europhys. Lett. {\bf 18},  457  (1992).

\bibitem{UE93a}
U. Eckern and A. Schmid, Ann. Phys. (Leipzig) {\bf 2},  180  (1993).

\bibitem{Smith92}
R.~A. Smith and V. Ambegaokar, Europhys. Lett. {\bf 20},  161  (1992).

\bibitem{Berkovits93}
M. Abraham and R. Berkovits, Phys. Rev. Lett. {\bf 70},  1509  (1993).

\bibitem{Mueller94}
A. M\"uller-Groeling and H.~A. Weidenm\"uller, Phys. Rev. B {\bf 49},  4752
  (1994).

\bibitem{Yoshioka94}
H. Kato and D. Yoshioka, Phys. Rev. B {\bf 50},  4943  (1994).

\bibitem{Giamarchi95}
T. Giamarchi and B.~S. Shastry, Phys. Rev. B {\bf 51},  10915  (1995).

\bibitem{Ramin95}
M. Ramin, B. Reulet, and H. Bouchiat, Phys. Rev. B {\bf 51},  5882  (1995).

\bibitem{Peters96}
P. Schmitteckert and U. Eckern, Phys. Rev. B {\bf 53},  1539  (1996).


\bibitem{UE96}
U. Eckern and P. Schwab, Adv. Phys. {\bf 44},  387  (1995).

\bibitem{Altshuler81a}
B.~L. Altshuler and A.~G. Aronov, Solid State Commun. {\bf 38},  11  (1981).

\bibitem{Altshuler85}
B.~L. Altshuler and A.~G. Aronov,  in {\em Electron-Electron Interactions in
  Disordered Systems}, edited by A.~L. Efros and M. Pollak (North-Holland,
  Amsterdam, 1985), p.\ 1.

\bibitem{UE93}
U. Eckern, Phys. Scr. T {\bf 49},  338  (1993).

\bibitem{Yoshioka93}
H. Yoshioka and H. Fukuyama, J. Phys. Soc. Jpn. {\bf 62},  612  (1993).

\bibitem{PS96a}
P. Schwab and U. Eckern, Ann. Phys. (Leipzig) {\bf 5},  57  (1996).

\bibitem{Hewson93}
A.~C. Hewson, {\em The Kondo Problem to Heavy Fermions} (Cambridge University
  Press, Cambridge, 1993).

\bibitem{Hewson93a}
A.~C. Hewson, Phys. Rev. Lett. {\bf 70},  4007  (1993).

\bibitem{Altshuler85a}
B.~L. Altshuler, A.~G. Aronov, D.~E. Khmelnitskii, and A.~F. Larkin,  in {\em
  Quantum Theory of Solids}, edited by I.~M. Lifshitz (MIR Publishers, Moscow,
  1980), p.\ 130.

\bibitem{Hikami80}
S. Hikami, A.~I. Larkin, and Y. Nagaoka, Prog. Theor. Phys. {\bf 63},  707
  (1980).

\end{thebibliography}
%
%
%
%
%

\newpage
{\center{\large\bf Figure Captions}}
\begin{description}
\item[Figure \ref{fig1}:] \refstepcounter{figure} \label{fig1}
One-cooperon contribution to the fluctuations in the local spin density. 
\item[Figure \ref{figDomegaSd}:] \refstepcounter{figure} \label{figDomegaSd}
Second harmonic of the grand potential normalized by the Thouless energy as a function
of temperature and Zeeman splitting. The spin-flip scattering rate is chosen equal 
to the Thouless energy; no spin-orbit scattering. 
\item[Figure \ref{figSpinOrb}:] \refstepcounter{figure} \label{figSpinOrb}
Second harmonic of the grand potential as a function
of temperature and Zeeman splitting. 
The spin-flip scattering rate and the Zeeman energy are the same as in Fig.\ \ref{figDomegaSd},
but here we consider the limit of strong spin-orbit scattering.
For comparison we also plot $\langle \Omega_2 \rangle$ due to the Coulomb interaction \cite{Ambeg90,UE91}.
Apparently, the impurity-spin induced contribution is dominant at low temperature.
\item[Figure \ref{figFeynhigh}:] \refstepcounter{figure} \label{figFeynhigh}
Higher order Feynman diagram. The curly line represents scattering at magnetic impurities.
\item[Figure \ref{figHighord}:] \refstepcounter{figure} \label{figHighord}
Second harmonic of the grand potential as a function of the spin-flip scattering rate.
\item[Figure \ref{figAnderson}:] \refstepcounter{figure} \label{figAnderson}
Graphical representation of the phase sensitive contribution to the grand potential in the
Anderson model. The double line represents the $d$-electron Green's function, the wavy line
is the on-site interaction $U$, and the cross denotes the hybridization $V$.                            
\item[Figure \ref{figCooperon}:] \refstepcounter{figure} \label{figCooperon}
Graphical representation of the equation defining the cooperon. Here, the dashed line represents
scattering at both magnetic and non-magnetic impurities.
\end{description}
\input{epsf}
\newpage\noindent
\epsfbox{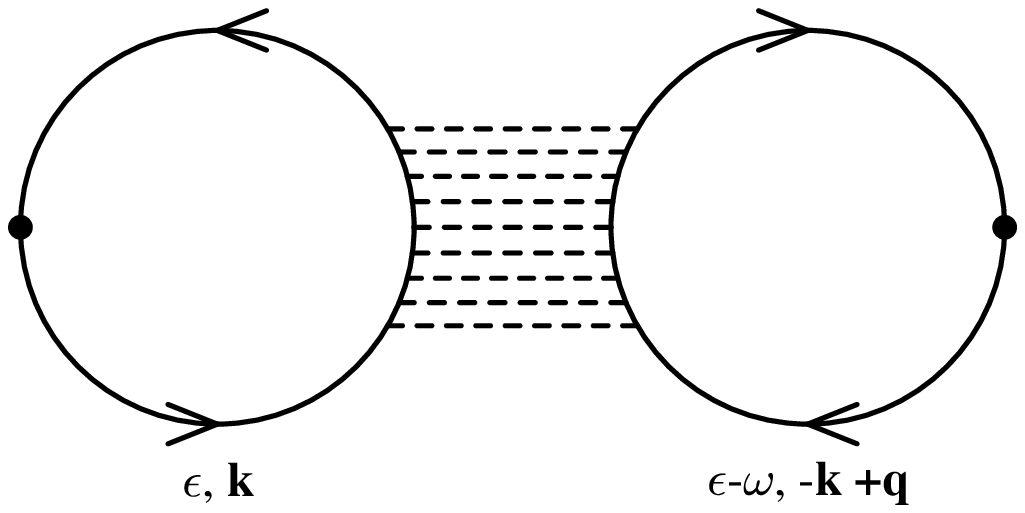}
\\[2ex]
\vfill Schwab et al.: Figure \ref{fig1}
\newpage\noindent

\setlength{\unitlength}{0.1bp}
\begin{picture}(3600,2160)(0,0)
\put(3054,1746){\makebox(0,0)[r]{$\omega_s = 5.0 E_c$}}
\put(3054,1846){\makebox(0,0)[r]{$\omega_s = 1.0 E_c$}}
\put(3054,1946){\makebox(0,0)[r]{$\omega_s = 0.1 E_c$}}
\put(2008,51){\makebox(0,0){$T/E_c$}}
\put(100,1180){%
\makebox(0,0)[b]{\shortstack{$\langle\Omega_2\rangle /E_c$}}%
}
\put(3417,151){\makebox(0,0){3}}
\put(2948,151){\makebox(0,0){2.5}}
\put(2478,151){\makebox(0,0){2}}
\put(2009,151){\makebox(0,0){1.5}}
\put(1539,151){\makebox(0,0){1}}
\put(1070,151){\makebox(0,0){0.5}}
\put(600,151){\makebox(0,0){0}}
\put(540,2109){\makebox(0,0)[r]{0}}
\put(540,1877){\makebox(0,0)[r]{-0.0005}}
\put(540,1645){\makebox(0,0)[r]{-0.001}}
\put(540,1412){\makebox(0,0)[r]{-0.0015}}
\put(540,1180){\makebox(0,0)[r]{-0.002}}
\put(540,948){\makebox(0,0)[r]{-0.0025}}
\put(540,716){\makebox(0,0)[r]{-0.003}}
\put(540,483){\makebox(0,0)[r]{-0.0035}}
\put(540,251){\makebox(0,0)[r]{-0.004}}
\end{picture}
\\[2ex]
\vfill Schwab et al.: Figure \ref{figDomegaSd}
\newpage\noindent

\setlength{\unitlength}{0.1bp}
\begin{picture}(3600,2160)(0,0)
\put(3054,1646){\makebox(0,0)[r]{Coulomb interaction}}
\put(3054,1746){\makebox(0,0)[r]{$\omega_s = 5.0 E_c$}}
\put(3054,1846){\makebox(0,0)[r]{$\omega_s = 1.0 E_c$}}
\put(3054,1946){\makebox(0,0)[r]{$\omega_s = 0.1 E_c$}}
\put(2008,51){\makebox(0,0){$T/E_c$}}
\put(100,1180){%
\makebox(0,0)[b]{\shortstack{$\langle \Omega_2 \rangle /E_c$}}%
}
\put(3417,151){\makebox(0,0){3}}
\put(2948,151){\makebox(0,0){2.5}}
\put(2478,151){\makebox(0,0){2}}
\put(2009,151){\makebox(0,0){1.5}}
\put(1539,151){\makebox(0,0){1}}
\put(1070,151){\makebox(0,0){0.5}}
\put(600,151){\makebox(0,0){0}}
\put(540,2109){\makebox(0,0)[r]{0.2}}
\put(540,1645){\makebox(0,0)[r]{0.15}}
\put(540,1180){\makebox(0,0)[r]{0.1}}
\put(540,716){\makebox(0,0)[r]{0.05}}
\put(540,251){\makebox(0,0)[r]{0}}
\end{picture}
\\[2ex]
\vfill Schwab et al.: Figure \ref{figSpinOrb}
\newpage\noindent
\epsfbox{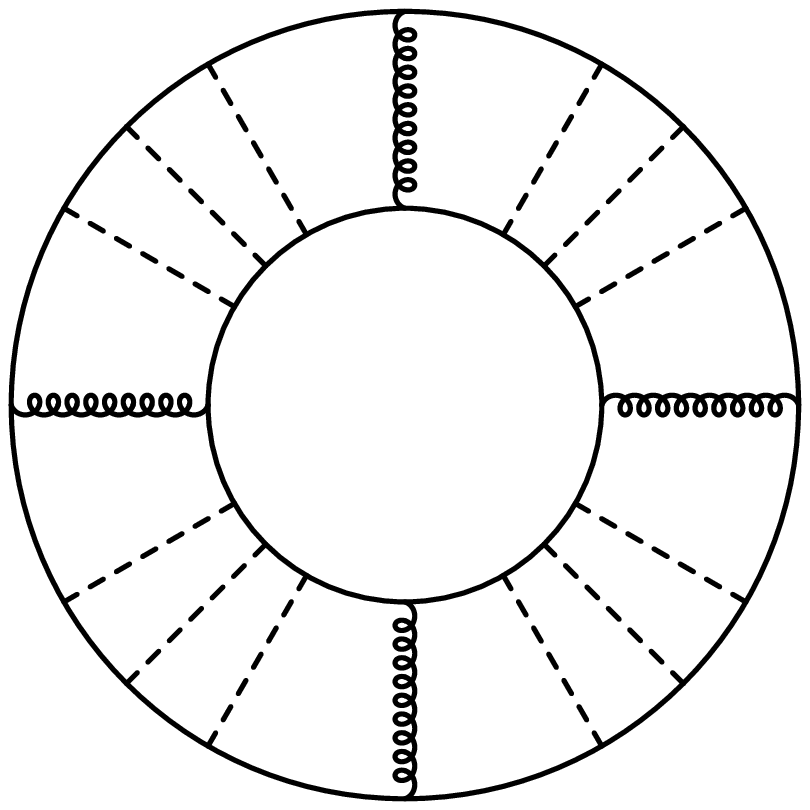}\\[2ex]
\vfill Schwab et al.: Figure \ref{figFeynhigh}
\newpage\noindent

\setlength{\unitlength}{0.1bp}
\begin{picture}(3600,2160)(0,0)
\put(2741,1390){\makebox(0,0)[r]{no spin-orbit scattering}}
\put(2741,1490){\makebox(0,0)[r]{strong spin-orbit scattering}}
\put(2008,51){\makebox(0,0){$1/\tau_s E_c$}}
\put(100,1180){%
\makebox(0,0)[b]{\shortstack{$T \langle \Omega_2 \rangle /E_c^2$}}%
}
\put(3417,151){\makebox(0,0){25}}
\put(2854,151){\makebox(0,0){20}}
\put(2290,151){\makebox(0,0){15}}
\put(1727,151){\makebox(0,0){10}}
\put(1163,151){\makebox(0,0){5}}
\put(600,151){\makebox(0,0){0}}
\put(540,2109){\makebox(0,0)[r]{0.15}}
\put(540,1799){\makebox(0,0)[r]{0.1}}
\put(540,1490){\makebox(0,0)[r]{0.05}}
\put(540,1180){\makebox(0,0)[r]{0}}
\put(540,870){\makebox(0,0)[r]{-0.05}}
\put(540,561){\makebox(0,0)[r]{-0.1}}
\put(540,251){\makebox(0,0)[r]{-0.15}}
\end{picture}
\\[2ex]
\vfill Schwab et al.: Figure \ref{figHighord}
\newpage\noindent
\epsfbox{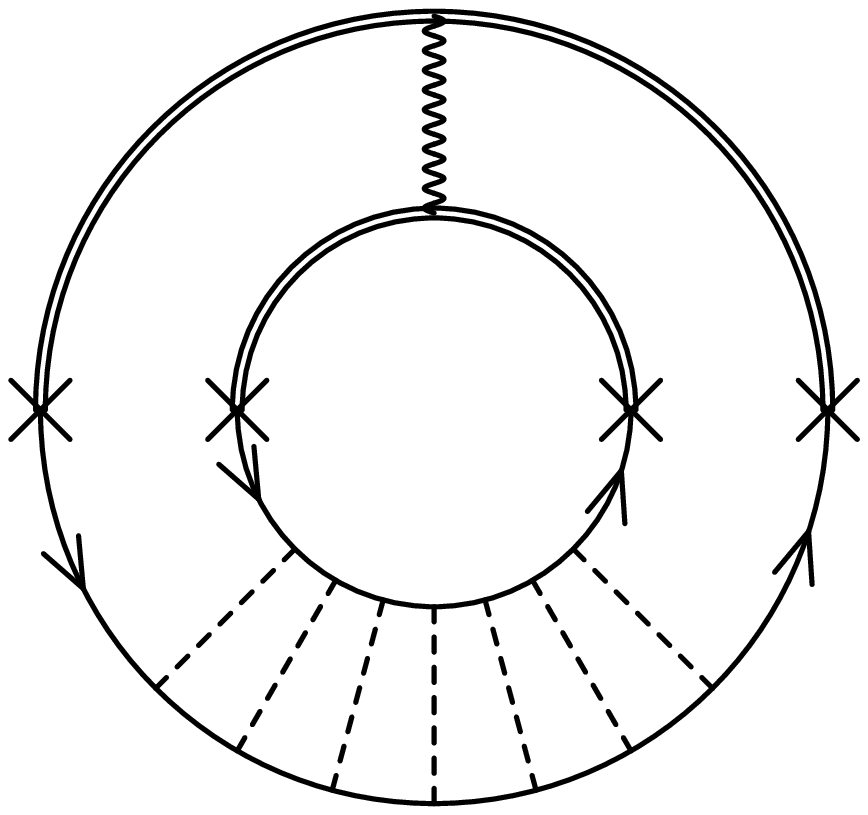}\\[2ex]
\vfill Schwab et al.: Figure \ref{figAnderson}
\newpage\noindent
\epsfbox{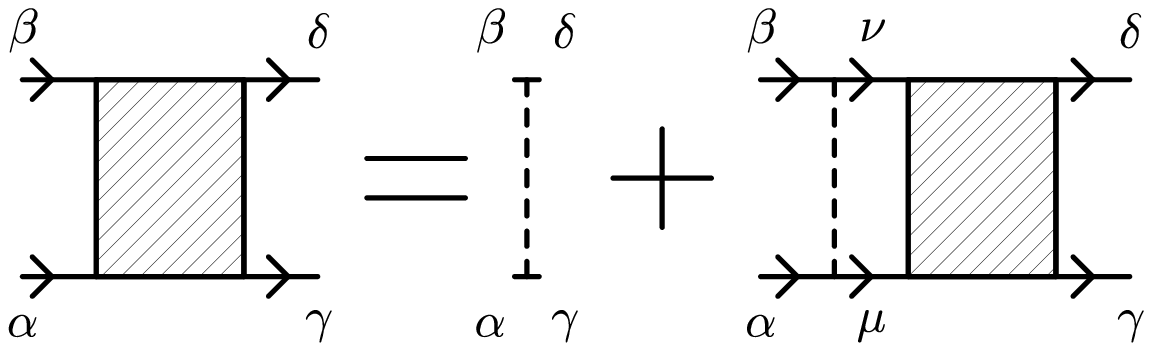}\\[2ex]
\vfill Schwab et al.: Figure \ref{figCooperon}
\end{document}